\documentclass[singlecolumn,epjc3]{svjour3}
\smartqed  
\RequirePackage{graphicx}
\journalname{Eur. Phys. J. C}

\begin{document}

\title{All spherically symmetric charged anisotropic solutions for compact stars}

\author{S.K. Maurya\thanksref{e1,addr1}
\and Y.K. Gupta\thanksref{e2,addr2}
\and Saibal Ray\thanksref{e3,addr3}.}

\thankstext{e1}{e-mail: sunil@unizwa.edu.om}
\thankstext{e2}{e-mail: kumar$001947$@gmail.com}
\thankstext{e3}{e-mail: saibal@associates.iucaa.in}

\institute{Department of Mathematical \& Physical Sciences, College of Arts \& Science,
University of Nizwa, Nizwa, Sultanate of Oman\label{addr1}
\and Department of Mathematics, Raj Kumar Goel Institute of Technology, Ghaziabad, 201003, U.P., India\label{addr2}
\and Department of Physics, Government College of Engineering \& Ceramic Technology, Kolkata 700010, West Bengal, India\label{addr3}}

\date{Received: date / Accepted: date}

\maketitle

\begin{abstract}
In the present paper we develop an algorithm for all spherically
symmetric anisotropic charged fluid distribution. Considering a
new source function $\nu(r)$ we find out a set of solutions which
is physically well behaved and represent compact stellar models. A
detailed study specifically shows that the models actually
correspond to strange stars in terms of their mass and radius. In
this connection we investigate about several physical properties
like energy conditions, stability, mass-radius ratio, electric
charge content, anisotropic nature and surface redshift through
graphical plots and mathematical calculations. All the features
from these studies are in excellent agreement with the already
available evidences in theory as well as observations.
\end{abstract}

\keywords{general relativity; electric field; anisotropic fluid; compact star}

\section{Introduction}

Historically the possibility that self gravitating stars could
actually contain a non-vanishing net charge was first pointed out
by Rosseland \cite{Rosseland1924} and later on by several other
researchers \cite{Neslusan2001,Anninos2001,Giuliani2008} with
different view points. The general relativistic analog for
charged dust stars were discussed by Majumdar \cite{Majumdar} and
Papapetrou \cite{Papapetrou}. However, in his pioneering work
Bonnor \cite{Bonnor} further discussed this issue and also several
investigators considered the problem later on in detailed in
connection to stability and other aspects
\cite{Bekenstein1971,Zhang1982,Felice1995a,Felice1999,Yu2000,Glendenning2000,Anninos2001,Ivanov2002}.

Following Treves and Turolla~\cite{Treves}, to justify the present
work with a charged fluid distribution, Ray and
Das~\cite{SaibalRay2002,SaibalRay2004} argue that even though the
astrophysical systems are by and large electrically neutral,
recent studies do not rule out the possibility of the existence of
massive astrophysical systems that are not electrically neutral.
The mechanism is mainly related to the acquiring a net charge by
accretion from the surrounding medium or even by a compact star
during its collapse from the supernova stage. In this connection
it is interesting to note that to study the effect of electric
charge in compact stars Ray et al. \cite{SubharthiRay2003} by
assuming an ansatz have shown that in order to see any appreciable
effect on the phenomenology of the compact stars, the total
electric charge is to be $\sim 10^{20}$ Coulomb.

It has been pointed out by Ivanov \cite{Ivanov2002} that
substantial analytical difficulties associated with
self-gravitating, static, isotropic fluid spheres when pressure
explicitly depends on matter density. However, it is also observed
that simplification can be achieved with the introduction of
electric charge. It is to note that charged, self-gravitating
anisotropic fluid spheres have been investigated by Horvat et al.
\cite{Horvat} in studies of gravastars and also recently
Thirukkanesh and Maharaj \cite{TM} found solutions for the charged
anisotropic fluid.

In connection to stability of the stellar model Stettner
\cite{Stettner} argued that a fluid sphere of uniform density with
a net surface charge is more stable than without charge.
Therefore, as pointed out by Rahaman et al. \cite{Rahaman} that a
general mechanism have been adopted to overcome singularity due to
gravitational collapsing of a static, spherically symmetric fluid
sphere is to include charge to the neutral system. It is observed
that in the presence of charge several features may arise: (i)
gravitational attraction is counter balanced by the electrical
repulsion in addition to the pressure gradient \cite{Sharma}, (ii)
it inhibits the growth of space-time curvature which has a great role
to avoid singularities \cite{Felice1995b} and (iii) the presence
of the charge function serves as a safety valve, which absorbs
much of the fine tuning, necessary in the uncharged case
\cite{Ivanov2002}.

One can notice that since the breakthrough idea of white dwarf by
Chandrasekhar \cite{Chandrasekhar1931} the study of compact stars
gained a tremendous motive in the field of ultra-dense objects. In
this line of research the other dense compact stars are neutron
stars, quark stars, strange stars, boson stars, gravastars and so
on. As far as composition is concerned in the compact stars the
matter is found to be in stable ground state where the quarks are
confined inside the hadrons. It is argued by several workers
\cite{Witten1984,Glendenning1997,Xu2003,DePaolis2007} that if it
is composed of the de-confined quarks then also a stable ground
state of matter, known as `strange matter', is achievable which
provides a `strange star'. There are two aspects of this
assumption behind strange star: (i) theoretically to explain the
exotic phenomena of gamma ray bursts and soft gamma ray repeaters
\cite{Cheng1996,Cheng1998}, and (ii) observationally confirmation
of $SAX~J$1808.4-3658 as one of the candidates for a strange star
by the Rossi X-ray Timing Explorer \cite{Li1999}.

It was Ruderman \cite{Ruderman1972} who investigated that the
nuclear matter may have anisotropic features at least in certain
very high density ranges ($> ~ 10^{15}~gm/cm^3$), where the
nuclear interaction must be treated relativistically. However,
later on Bowers and Liang \cite{Bowers1974} showed specifically
that anisotropy might have non-negligible effects on such
parameters like maximum equilibrium mass and surface redshift. We
notice that recently anisotropic matter distribution has been
considered by several authors in connection to compact stars
~\cite{Mak2002,Mak2003,Usov2004,Varela2010,Rahaman2010a,Rahaman2011,Rahaman2012a,Kalam2012}.

Studies have been shown that at the centre of the fluid sphere the
anisotropy vanishes. However, for small radial increase the
anisotropy parameter increases, and after reaching a maximum in
the interior of the star, it becomes a decreasing function of the
radial distance \cite{Mak-Harko}. So there are several
possibilities of expressions for charge functions and pressure
anisotropy. It is also indicated by Varela et al.
\cite{Varela2010} that inward-directed fluid forces caused by
pressure anisotropy may allow equilibrium configurations with
larger net charges and electric field intensities than those found
in studies of charged isotropic fluids.

Algorithm for perfect fluid and anisotropic uncharged fluid is
already published by others \cite{Lake2003,Lake2004,Herrera2008}.
In his work Lake \cite{Lake2003,Lake2004} has considered an
algorithm based on the choice of a single monotone function which
generates all regular static spherically symmetric perfect as well
as anisotropic fluid solutions of Einstein's equations. On the
other hand, Herrera et al. \cite{Herrera2008} have extended the
algorithm to the case of locally anisotropic fluids. Therefore,
there remains a natural choice of an algorithm to a more general
case with the inclusion of charge along with anisotropic fluid
distribution.

Under the above background and motivation, therefore in the
present paper, we have carried out investigation for a
relativistic stellar model with charged anisotropic fluid sphere.
The schematic format of this study is as follows: We provide the
Einstein-Maxwell field equations for charged anisotropic stellar
source in Sect. 2 whereas allied algorithm has been constructed in
Sect. 3. The general solutions are shown in Sect. 4, along with a
special example for the index $n=1$ and matching of the interior
solution with the exterior Reissner-Nordstr{\"o}m solution. In
Sect. 5 we explore several interesting properties of the physical
parameters which include density, pressure, stability, charge,
anisotropy and redshift. Special case studies have been conducted
in Sect. 6 to verify (i) mass-radius ratio and (ii) density of the
star both of which clearly indicate that the model represents
stable configuration of a strange compact star. Sect. 7 is devoted
as a platform for providing some salient features and concluding
remarks.

\section{The Field Equations for Charged and Anisotropic Matter Distribution}

In this work we intend to study a static and spherically symmetric
matter distribution whose interior metric is given in
Schwarzschild coordinates \cite{Tolman1939,Oppenheimer1939}
$x^i=(r, \theta, \phi, t)$ as follows:
\begin{equation}
ds^2 = - e^{\lambda(r)} dr^2 - r^2(d\theta^2 + sin^2\theta d\phi^2) + e^{\nu(r)} dt^2.\label{metric1}
\end{equation}

The Einstein-Maxwell field equations are as usual given by
\begin{equation}
 {R^i}_j - \frac{1}{2} R\, {g^i}_j = \kappa ({T^i}_j + {E^i}_j), \label{field}
\end{equation}
where $\kappa = 8\pi$ is the Einstein constant with $G=1=c$ in
relativistic geometrized unit, $G$ and $c$ respectively being the
Newtonian gravitational constant and velocity of photon in vacua.

The matter within the star is assumed to be locally anisotropic
fluid in nature and consequently ${T^i}_j$ and ${E^i}_j$ are the
energy-momentum tensor of fluid distribution and electromagnetic
field defined by \cite{Dionysiou1982}
\begin{equation}
{T^i}_j = [(\rho + p_t)v^iv_j - p_t{\delta^i}_j + (p_r - p_t) \theta^i \theta_j],\label{matter}
\end{equation}

\begin{equation}
{E^i}_j = \frac{1}{4}(-F^{im}F_{jm} + \frac{1}{4}{\delta^i}_jF^{mn}F_{mn}),\label{electric}
\end{equation}
where $v^i$ is the four-velocity as
$v^i=e^{\nu(r)/2}{\delta^i}_4$, $\theta^i$ is the unit space like
vector in the direction of radial vector as $\theta^i =
e^{\lambda(r)/2}{\delta^i}_1$, $\rho $ is the energy density,
$p_r$ is the pressure in the direction of $\theta^i$ (normal or
radial pressure) and $p_t$ is the pressure orthogonal to
$\theta_i$ (transverse or tangential pressure), while ${T^1}_1=-p_r$,\, ${T^2}_2={T^3}_3=-p_t$,\, ${T^4}_4=\rho$
and ${E^1}_1=-{E^2}_2=-{E^3}_3={E^4}_4=\frac{1}{8\,\pi}\,\frac{q^2(r)}{r^4}$.\\

Now, anti-symmetric electromagnetic field tensor $F_{ij}$ can be
defined by
\begin{equation}
F_{ij} = \frac{\partial A_j}{\partial x_i} - \frac{\partial A_i}{\partial x_j},
\end{equation}
which satisfies the Maxwell equations
\begin{equation}
F_{ik,j} + F_{kj,i} + F_{ji,k} = 0,
\end{equation}

\begin{equation}
\frac{\partial}{\partial x^k}({\sqrt -g} F^{ik}) = -4\pi{\sqrt -g}
J^i, \label{max2}
\end{equation}
where $g$ is the determinant of quantities $g_{ij}$ in Eq.
(\ref{field}) defined by
\begin{equation}
g = \left(\begin{array}{cccc} e^{\nu} & 0           & 0    & 0\\
                              0       &-e^{\lambda} & 0    & 0\\
                              0       & 0           & -r^2 & 0\\
                              0       & 0           & 0    & -r^2sin^2\theta \end{array} \right) = - e^{\nu+\lambda}r^4 sin^2\theta,
\end{equation}
where, $A_j=(\phi(r), 0, 0, 0)$  is four-potential and $J^i$ is
the four-current vector defined by
\begin{equation}
J^i = \frac{\sigma}{\sqrt g_{44}} \frac{dx^i}{dx^4} = \sigma v^i,\label{J11}
\end{equation}
where $\sigma$ is the charged density.

For static matter distribution the only non-zero component of the
four-current is $J^4$. Because of spherical symmetry, the
four-current component is only a function of radial distance, $r$.
The only non vanishing components of electromagnetic field tensor
are $F^{41}$ and $F^{14}$, related by $F^{41} = - F^{14}$, which
describe the radial component of the electric field. From the Eq.
(\ref{max2}) and (\ref{J11}), one obtains the following expression for the component of
electric field:
\begin{equation}
e^{(\nu+\lambda)/2}\,r^2\,F^{41} =-4\pi\,\int_0^r {\frac{\sigma\, r^2\, e^{(\lambda+\nu)/2}}{\sqrt{g_{44}}}\, dr} =-4\pi\,\int_0^r {{\sigma r^2 e^{\lambda/2}}\,dr},
\label{F14}
\end{equation}
where $\sqrt{g_{44}}=e^{\nu/2}$ and  if $q(r)$ represents the total charge contained within the
sphere of radius $r$, then it can be defined by the relativistic Gauss law as
\begin{equation}
q(r) = 4\pi \int_0^r \sigma r^2 e^{\lambda/2} dr = r^2 \sqrt{-F_{14}F^{14}}. \label{charge}
\end{equation}

From Eqs. (\ref{F14}) and (\ref{charge}), we obtain the electric charge $q(r)$ as
\begin{equation}
q(r)=- e^{(\nu+\lambda)/2}\,r^2\,F^{41}.
\end{equation}

For the spherically symmetric metric (\ref{metric1}), the
Einstein-Maxwell field equations may be expressed as the following
system of ordinary differential equations \cite{Dionysiou1982}
\begin{equation}
-\kappa ({T^1}_1+{E^1}_1) = \frac{{\nu}^{\prime}}{r} e^{-\lambda} - \frac{(1 - e^{-\lambda})}{r^2} = \kappa p_r - \frac{q^2}{r^4},\label{e1}
\end{equation}

\begin{eqnarray}
-\kappa ({T^2_2}+{E^2}_2) = -\kappa ({T^3}_3+{E^3}_3)= \left[\frac{{\nu}^{\prime\prime}}{2} - \frac{{\lambda}^{\prime}{\nu}^{\prime}}{4} + \frac{{{\nu}^{\prime}}^2}{4} + \frac{{\nu}^{\prime} - {\lambda}^{\prime}}{2r}\right]e^{-\lambda} \nonumber\\ = \kappa p_t + \frac{q^2}{r^4},\label{e2}
\end{eqnarray}

\begin{equation}
\kappa ({T^4}_4+{E^4}_4) = \frac{{\lambda}^{\prime}}{r} e^{-\lambda} + \frac{(1 - e^{-\lambda})}{r^2} = \kappa \rho + \frac{q^2}{r^4},\label{e3}
\end{equation}
where the prime denotes differential with respect to $r$.

If the mass function for electrically charged fluid sphere is
denoted by $m(r)$, then it can be defined by the metric function
$e^{\lambda(r)}$ as
\begin{equation}
e^{-\lambda(r)} = 1 - \frac{2m(r)}{r} + \frac{q^2}{r^2}.\label{e4}
\end{equation}

If $R$ represents the radius of the fluid spheres then it can be
showed that $m$ is constant $m(r = R) = M$ outside the fluid
distribution where $M$ is the gravitational mass. Thus the
function $m(r)$ represents the gravitational mass of the matter
contained in a sphere of radius $r$. The gravitational mass $M$ of
the fluid distribution is defined as
\begin{equation}
M = \mu(R) + \xi(R),\label{mass}
\end{equation}
where $\mu(R) = \frac{\kappa}{2} \int_0^R\rho\, r^2\, dr$ is the mass
inside the sphere, $\xi(R) = \frac{\kappa}{2} \int_0^R \sigma\, r\, q\,
e^{\lambda/2}\,dr$ is the mass equivalence of the electromagnetic
energy of distribution and $q(R)$ is the total charge inside
the fluid spheres \cite{Florides1983}.

Now using Eq. (\ref{mass}) and Eq. (11), we can write the mass
$m(r)$ of the fluid spheres of radius $r$ in terms of energy
density and charge function as
\begin{equation}
m(r)= \frac{\kappa}{2} \int \rho r^2 dr +\frac{1}{2} \int
\frac{q^2}{r^2}dr + \frac{q^2}{2r},\label{e5}
\end{equation}
whereas from Eqs. (\ref{e1}) and (\ref{e4}) we obtain
\begin{equation}
\nu^{\prime} = \frac{\left(\kappa r p_r + \frac{2m}{r^2} - \frac{2q^2}{r^3}\right)}{\left(1 - \frac{2m}{r} + \frac{q^2}{r^2}\right)}.\label{e6}
\end{equation}

We suppose here that the radial pressure is not equal to the
tangential pressure i.e. $p_r \neq p_t$, otherwise if the radial
pressure is equal to the transverse pressure i.e. $p_r = p_t$,
which corresponds to isotropic or perfect fluid distribution. Let
the measure of anisotropy $\Delta = p_t - p_r$ and is called the
anisotropy factor \cite{Herrera1985}. The term $2(p_t - p_r)/r$
appears in the conservation equations ${T^i}_{j;i} = 0$ (where,
semi-colon denotes the covariant derivative) which is representing
a force due to anisotropic nature of the fluid. When $p_t > p_r$
then direction of force to be outward and inward when $p_t < p_r$.
However, if $p_t > p_r$, then the force allows construction of
more compact object for the case of anisotropic fluid than
isotropic fluid distribution \cite{Gokhroo1994}.

By using Eqs. (\ref{e1})-(\ref{e4}) and also Eqs. (\ref{e5}) and
(\ref{e6}) the expression of pressure gradient in terms of mass,
charge, energy density and radial pressure read as
\begin{equation}
\frac{dp_r}{dr} =- \frac{\left(p_r + \frac{2m^{\prime}}{r^2} - \frac{q}{r^3}\frac{dq}{dr}\right)\left(\kappa r p_r + \frac{2m}{r^2}
- \frac{2q^2}{r^3}\right)}{2\left(1 - \frac{2m}{r} + \frac{q^2}{r^2}\right)}+\frac{q}{4\pi r^4} \frac{dq}{dr} + \frac{2\Delta}{r},\label{e7}
\end{equation}
where $m^{\prime} \equiv \frac{dm}{dr}$ i.e. variation of mass
with radial coordinate $r$. The above Eq.~(\ref{e7}) represents
the charged generalization of the well-known
Tolman-Oppenheimer-Volkoff (TOV) equation of hydrostatic for
anisotropic stellar structure \cite{Tolman1939,Oppenheimer1939}.

\section{The Algorithm for Constructing all Possible Anisotropic Charged Fluid Solutions}

The Einstein equations Eqs. (\ref{e1}), (\ref{e2}) and (\ref{e3})
in terms of mass function reduce to as follows:
\begin{equation}
-\frac{2m(1+r{\nu}^{\prime})}{r^3} + \frac{{\nu}^{\prime}}{r}+\frac{q^2(1+r{\nu}^{\prime})}{r^4}+ \frac{q^2}{r^4}=\kappa p_r,\label{e8}
\end{equation}

\begin{eqnarray}
-\frac{m^{\prime}(2+r{\nu}^{\prime})}{2r^2} - \frac{m(2r^2{\nu}^{\prime\prime} +r^2{{\nu}^{\prime}}^2 +r{\nu}^{\prime} -2)}{2r^3} \nonumber\\
+ \frac{2rqq^{\prime}{\nu}^{\prime}-2q^2{\nu}^{\prime}+4qq^{\prime} +(r^2+q^2)(2r{\nu}^{\prime\prime} +r{{\nu}^{\prime}}^2+2{\nu}^{\prime})}{4r^3}
- \frac{2q^2}{r^4}=\kappa p_t,\label{e9}
\end{eqnarray}

\begin{equation}
\frac{2m^{\prime}}{r^2} - \frac{2qq^{\prime}}{r^3}= \kappa \rho.\label{e10}
\end{equation}

Using Eqs. (\ref{e8}) and (\ref{e9}), we obtain a Riccati equation
in the first derivative of $\nu(r)$. However, after the re
examination of the differential equation we come across a linear
differential equation of first order in $m(r)$ \cite{Berger1987}.

The first order linear differential equation of $m(r)$ in terms of
$\nu(r)$, anisotropy $\Delta = (p_t - p_r)$ and charge function
$q(r)$ can be provided as follows:
\begin{equation}
m^{\prime}+ \frac{(2r^2{\nu}^{\prime\prime} + r^2{{\nu}^{\prime}}^2 - 3r{\nu}^{\prime} -6)}{r(r{\nu}^{\prime} + 2)}m
= \frac{(2r^2{\nu}^{\prime\prime} + r^2{{\nu}^{\prime}}^2 - 2{\nu}^{\prime}r)}{2(r{\nu}^{\prime} + 2)} + f(r),\label{e11}
\end{equation}
where
\begin{equation}
f(r)= \frac{2rq^2{\nu}^{\prime\prime} + q^2{\nu}^{\prime}(r{\nu}^{\prime} - 4) + 2qq^{\prime}(r{\nu}^{\prime} + 2)}{2r(r{\nu}^{\prime} + 2)}
- \frac{2r^2}{(r{\nu}^{\prime} + 2)} \left(\Delta + \frac{4q^2}{r^4}\right).\label{e12}
\end{equation}

The above Eq. (\ref{e11}) gives the mass $m(r)$ as follows:
\begin{equation}
m(r)= e^{-\int g(r) dr} \left[\int\left\{h(r) + f(r)\right\}\left(e^{\int g(r)dr}\right) dr + A \right],\label{e13}
\end{equation}
where
\begin{equation}
g(r) = \frac{(2r^2{\nu}^{\prime\prime} + r^2{{\nu}^{\prime}}^2 - 3r{\nu}^{\prime} -6)}{r(r{\nu}^{\prime} + 2)},\label{e14}
\end{equation}

\begin{equation}
h(r) = \frac{(2r^2{\nu}^{\prime\prime} + r^2{{\nu}^{\prime}}^2 - 2{\nu}^{\prime}r)}{2(r{\nu}^{\prime} + 2)},\label{e15}
\end{equation}
where we have used the symbol $\prime \equiv \frac{d}{dr}$.

At this point we would like to construct useful algorithm to
generate solutions for any known generic function $\nu(r)$. Now
from Eqs. (\ref{e8}) and (\ref{e10}), we get
\begin{equation}
\kappa \rho = \frac{2m^{\prime}}{r^2} - \frac{2qq^{\prime}}{r^3} \geq 0,\label{e16}
\end{equation}

\begin{equation}
\kappa p_r = \frac{r[\nu^{\prime}(r^2+q^2 - 2rm) - 2m] + 2q^2}{r^4} \geq 0.\label{e17}
\end{equation}

Note that the inequalities in (\ref{e16}) and (\ref{e17}) are to
be viewed from reality or energy conditions which will impose the restrictions on $\nu(r)$. At the centre of
symmetry $(r = 0)$ the regularity of the Ricci invariants requires
that energy density $\rho(r)$, radial pressure $p_r(r)$ and
tangential pressure $p_t(r)$ at origin should be finite. The
regularity of Weyl invariants requires that mass $m(r)$ and charge
$q(r)$ at $r=0$ should satisfy: $m(0)=
m^{\prime}(0)=m^{\prime\prime}(0)=0$, $q(0)= q^{\prime}(0)=0$ and
$m^{\prime\prime\prime}(0)=\kappa\rho(0) +
(q^{\prime\prime}(0))^2$.

Now the metric function $\nu(0)$ is a finite constant, $q(0)= 0$
and it follows from (\ref{e17}) that $\nu^{\prime}(0)=0$ and
$\nu^{\prime\prime}(0)=\frac{\kappa}{3}[\rho(0)+3p_r(0)]-(q^{\prime\prime}(0))^2
> 0$. Since $\rho \geq 0$ and continuous, and also since $p_r > 0$
and finite, therefore it follows that $r > 2m(r)$
\cite{Baumgarte1993,Mars1996}. With $r > 2m(r)$ for $r > 0$. It
also follows from (\ref{e17}) for $p_r > 0$ that
$\nu^{\prime}(r)\neq 0$. As a result, the source function $\nu(r)$
must be a monotone increasing function with a regular minimum at
$r = 0$.

\section{A Class of New Solutions for Charged Anisotropic Stellar Models}
For a class of new anisotropic charged stellar models we consider
the following suitable source function in the form of metric
potential as follows:
\begin{equation}
\nu(r) = -n~log~B^{-1/n} (1-Cr^2),\label{e18}
\end{equation}
where $n$, $B$ and $C$ are positive integers. It is suitable in
the sense that the source function given by Eq.~(\ref{e18}) is
monotonic increasing with a regular minimum at $r=0$. It is to
note that charged and uncharged perfect fluid of this source
function with different electric intensity has already been
carried out \cite{Maurya2012a,Maurya2014a} where it was proved
that the above kind of source function with increasing and
non-singular behaviour provides physically valid solutions.

In terms of the source function expressed in Eq.~(\ref{e18}) we
consider the electric charge distribution and anisotropic pressure
distribution are in the following forms:
\begin{equation}
\frac{2q^2}{Cr^4} = K(Cr^2)^{1+N}[1+(n-1)Cr^2]^{\frac{n+1}{n-1}}(1-Cr^2)^{n+1}(a-bCr^2)^m,\label{e19}
\end{equation}

\begin{equation}
\frac{\Delta}{C} = \beta(Cr^2)^{n}[1+(n-1)Cr^2]^{n-1},\label{e20}
\end{equation}
where $K$, $N$ and $\beta$ are positive constants, $a$ and $b$ are
positive real numbers and $m$ is a positive integer. The electric
field intensity and anisotropy are vanishing at the center and
remains continuous, regular and bounded in the inside of the fluid
sphere for certain range of values of the parameters. Also these
forms of electric intensity and anisotropy function allow us to
integrate Eq. (\ref{e13}). Thus these choices may be physically
reasonable and useful in the study of the gravitational behavior
of anisotropic charged stellar models.

It is observed that Durgapal and Pandey \cite{Durgapal1984}, Ishak
et al. \cite{Ishak2001}, Lake \cite{Lake2003}, Pant
\cite{Pant2011} and Maurya et al. \cite{Maurya2012b} have proposed
solutions via the {\it ansatz} (\ref{e18}) with some particular
values of $n$. After that Maurya and Gupta
\cite{Maurya2011,Maurya2012c} showed that the same ansatz
for the metric function (1) by taking $n$ is a negative integer,
$C<0$ and $C>0$, $0<n<1$ and it produces an infinite family of
analytic solutions of the self-bound type (see details in the
Tables 7 and 8 of {\bf Appendix}). Recently Maurya and Gupta
\cite{Maurya2013} have also obtained infinite family of
anisotropic solutions for the same ansatz. But recently Murad
\cite{Murad2014} obtained charged stellar model for $n=-2$ and
$C<0$, however neutral solutions of this are irregular in the
behaviour of $dp/d\rho$ (Durgapal and Fuloria \cite{Durgapal1985},
Delgaty and Lake \cite{Delgaty1998}, Pant \cite{Pant2011}, Maurya
and Gupta \cite{Maurya2012c}). Hence the solution is not suitable
for application to a neutron star model because the equations of
state for nuclear matter show a regular behavior of $dp/d\rho$
\cite{Durgapal1985}. So in the present problem we have started
with regular behavior of $dp/d\rho$ in the same ansatz by taking
the value of $n=1$ and $3$. Recently Maurya et al.
\cite{Maurya2014b} argued that neutral solutions for these cases
have the regular behavior of  $dp/d\rho$ and it may be suitable
for application to a neutron star model.

By using together Eqs. (\ref{e18}), (\ref{e19}) and (\ref{e20}), the Eq. (\ref{e13}) gives $m(r)$ in the following form:
\begin{equation}
m(r)= e^{-\int G(r) dr} \left[\int\left\{H(r) + F(r)\right\}\left(e^{\int G(r)dr}\right) dr + A \right],\label{e21}
\end{equation}
where
\begin{equation}
G(r) = \frac{2(2n^2 + 5n - 3)C^2r^4 + (12-n)Cr^2 - 6}{2r(1 - Cr^2)[1+(n-1)Cr^2]},\label{e22}
\end{equation}

\begin{equation}
H(r) = \frac{n(n + 2)C^2r^4}{(1 - Cr^2)[1+(n-1)Cr^2]},\label{e23}
\end{equation}

\begin{equation}
F(r)= \frac{2K(Cr^2)^{N+2}(1 - Cr^2)^{n+1}(a-bCr^2)^m [F_1(r)+
[1+(n-1)Cr^2]F_2(r)]}{4(1 - Cr^2)[1+(n-1)Cr^2]} -
F_3(r),\label{e24}
\end{equation}
where
\begin{equation}
F_1(r)= [1+(n-1)Cr^2]^{\frac{n+1}{n-1}}[4nC^2r^4 - 2nCr^2],\label{e25}
\end{equation}

\begin{equation}
F_2(r)= 1 - 2Cr^2 + (2+N)(1 - Cr^2) - \frac{bmCr^2(1-Cr^2)}{(a-bCr^2)} + \frac{(1+n)Cr^2(1 - Cr^2)}{[1+(n-1)Cr^2]},\label{e26}
\end{equation}

\begin{eqnarray}
F_3(r)=
\frac{\beta(Cr^2)^{n+1}[1+(n-1)Cr^2]^{n-1}}{(1 - Cr^2)^{-n-1}[1+(n-1)Cr^2]} \nonumber\\
+ \frac{2K(Cr^2)^{N+2}[1+(n-1)Cr^2]^{\frac{n+1}{n-1}}(1-Cr^2)(a-bCr^2)}{(1 - Cr^2)^{-n-1}[1+(n-1)Cr^2]}.\label{e27}
\end{eqnarray}

In the absence of electric field intensity $(K =0)$ and pressure
anisotropy $(\beta= 0)$, the Eqs. (\ref{e8}), (\ref{e9}) and
(\ref{e10}) reduce to the equations obtained by Maurya et al.
\cite{Maurya2014b}. Corresponding solutions belongs to the
solutions of Maurya et al.
\cite{Maurya2011,Maurya2012c,Maurya2014b} for the values of $n$
as: all negative integers, all positive fractional values between
$0$ and $1$ and some positive integers ($n=1$, $2$ ~and ~$3$) and
solutions for particular values of $n$ to the well known Tolman
\cite{Tolman1939} for $n =-1$, Wyman \cite{Wyman1949}, Kuchowicz
\cite{Kuchowicz1975}, Adler \cite{Adler1974}, Adams and Cohen
\cite{Adams1975} all for $n =- 2$, Heintzmann
\cite{Heintzmann1969} for $n =-3$, Durgapal \cite{Durgapal1982}
for $n = -4, - 5$ and Pant \cite{Pant2011} for $n = -6, - 7$
for the {\it ansatz} (\ref{e18}).

\subsection{An Example: Physical parameters of Charged Anisotropic Model for $n=1$}

We calculate mass of the charged anisotropic fluid sphere as
\begin{eqnarray}
\frac{2m_1(r)}{r} = 1 - K(1 - Cr^2)^2 e^{2Cr^2}
[-\frac{1}{2}(Cr^2)^{N+2}(a-bCr^2)^m + Cr^2(1 - Cr^2)
\Pi_{a,b,C,m,i,N}(r)] \nonumber\\ + 2(\beta+3)Cr^2(1 -
Cr^2)^3e^{2Cr^2-2}[\phi_{C,j}(r) + \psi_{C,j}(r)] \nonumber\\- A(1
- Cr^2)^3 Cr^2 e^{2Cr^2} - [1+(\beta-2)Cr^2](1 -
Cr^2)^2,~~~~~~~~~~~~~~\label{e28}
\end{eqnarray}
where\\
$\phi_{C,j}(r) = \Sigma_{j=1}^{\infty}(-1)^{j-1} \frac{(1 - 2Cr^2)^j}{j}=log(2-2Cr^2)$,\\
$\psi_{C,j}(r) = \Sigma_{j=1}^{\infty}\frac{2(1 - Cr^2)^j}{j!j}=Ei(2-2Cr^2)-log(2-2Cr^2)$,\\
$\Pi_{a,b,C,m,i,N}(r) = \Sigma_{i=0}^{m}(-1)^i a^{m-i} b^i
\left(\begin{array}{c} m\\i \end{array}\right) \left[\frac{(Cr^2)^{N+i+1}}{N+i+1}\right]$.

The expressions for energy density, radial pressure and tangential
pressure are (by taking $x=Cr^2$) given by
\begin{eqnarray}
\frac{\kappa\rho(r)}{C} = A(4x^3-3x^3-7x^2-1)e^{2x} + (6-11x^2+2x^3) - \beta(3-10x+7x^2) \nonumber\\
+ (6+\beta)(1-x)^2(3-5x-4x^2)e^{2x-2}[\phi_{j}(x) + \psi_{j}(x)]\nonumber\\
-2\beta x e^{2x-2} (1-x)^2[1+(1-x)\theta_j(x)]\nonumber\\-K(1-x)^2(3-5x-4x^2)e^{2x} \Pi_{a,b,C,m,i,N}(x) \nonumber\\
-\frac{K}{2}(x)^{N+1}(1 - x)(a-bx)^m[4e^{2x} + e^{2x}(1-x)],\label{e29}
\end{eqnarray}

\begin{eqnarray}
\frac{\kappa p_r}{C} = A(1-x-x^2-x^3)e^{2x} + (2-10x+7x^2-2x^3)+
\frac{K}{2}e^{2x}(x)^{N+1}(1 - x)^2(a-bx)^m \nonumber\\- 2(6+\beta)(1+x-5x^2+3x^3)e^{2x-2}[\phi_{j}(x)
+ \psi_{j}(x)]+\beta(1-x^2)\nonumber\\+\frac{K(1-6x^2+8x^3-3x^4)e^{2x}}{(1 + x)} \Pi_{a,b,C,m,i,N}(x)],~~~~\label{e30}
\end{eqnarray}

\begin{eqnarray}
p_t = p_r + \Delta,\label{e31}
\end{eqnarray}
where $\Delta$ is the measure of anisotropy as defined earlier and
also\\ $\phi_{j}(x) = \sum_{j=1}^{\infty}(-1)^{j-1} \frac{(1 -
2x)^j}{j}$,\\ $\psi_{j}(x) = \sum_{j=1}^{\infty}\frac{2(1 -
x)^j}{j!j}$,\\ $\Pi_{a,b,C,m,i,N}(x) = \sum_{i=0}^{m}(-1)^i
a^{m-i} b^i \left(\begin{array}{c} m\\
                              i \end{array}\right) \left[\frac{(x)^{N+i+1}}{N+i+1}\right]$,\\
$\theta_{j}(x) = 2\sum_{j=1}^{\infty}\frac{2(1 - x)^{j-1}}{j!}$.

In a similar way one can calculate the mass $m(r)$ of charged
anisotropic model for $n=3$ and other permissible cases.

\subsection{Matching and Boundary Conditions}
The metric or first fundamental form of the boundary surface
should be the same whether obtained from the interior or exterior
metric, guarantees that for some coordinate system the metric
components $g_{ij}$ will be continuous across the surface. The
requirements of matching condition for  metric (1) that the above
system of equations is to be solved subject to the boundary
condition that radial pressure $p_r =0$ at $r=R$ (which is the
outer boundary of the fluid sphere). It is clear that $m (r=R)=M$
is a constant and, in fact, the interior metric (1) can be joined
smoothly at the surface of spheres $(r=R)$ to an exterior
Reissner-Nordstr{\"o}m metric whose mass is same as $m(r=R)=M$
\cite{Misner1964}. Thus one can get
\begin{equation}
ds^2 = - \left(1 - \frac{2M}{r} + \frac{Q^2}{r^2}\right)^{-1} dr^2
- r^2(d\theta^2 + sin^2\theta d\phi^2) + \left(1 - \frac{2M}{r} +
\frac{Q^2}{r^2}\right) dt^2,\label{metric2}
\end{equation}
which requires the continuity of $e^{\lambda(r)}$, $e^{\nu(r)}$
and $q$ across the boundary $r=R$
\begin{equation}
e^{-\lambda(R)}=e^{\nu(R)}= 1 - \frac{2M}{R} + \frac{Q^2}{R^2},
\end{equation}

\begin{equation}
q(R)=Q,
\end{equation}
where $M$ and $Q$ are called the total mass and charge inside the
fluid sphere respectively.

The continuity of $e^{\lambda(r)}$ and $e^{\nu(r)}$ on the
boundary is $e^{-\lambda(R)}=e^{\nu(R)}$, which gives the constant
$B$ in the following form:
\begin{equation}
B = (1-X)^n e^{-\lambda(R)} = (1-X)^n \left(1 - \frac{2M}{R} +
\frac{Q^2}{R^2}\right),\label{e47}
\end{equation}
where $X = CR^2$.

On the other hand, the arbitrary constant $A$ will be determined
from the boundary conditions by putting radial pressure $p_r =0$
at $r = R$ for the case $n=1$ as follows:
\begin{eqnarray}
A = \frac{-e^{2X}}{(1 - X)^2(1 + X)}\left[(2-10X+7X^2-2X^3)+
\frac{K}{2}e^{2X}(X)^{N+1}(1 - X)^2(a-bX)^m \right. \nonumber
\\ \left. - 2(6+\beta)(1+X-5X^2+3X^3)e^{2X-2}[\phi_{j}(X) + \psi_{j}(X)] +
\beta(1-X^2)\right. \nonumber
\\ \left. + \frac{K(1-6X^2+8X^3-3X^4)e^{2X}}{(1 + X)}
\Pi_{a,b,C,m,i,N}(X)\right].\label{e48}
\end{eqnarray}

Hence the total charge inside the star, central density and
surface density can respectively be evaluated for the case $n=1$
as follows:
\begin{equation}
Q (q_{r=R}) = R \sqrt{\frac{K}{2}e^{2X}(X)^{N+2}(1 -
X)^2(a-bX)^m}.\label{e30}
\end{equation}

\begin{equation}
\rho_0 = \frac{C}{\kappa}[-A +6
-3\beta+3(6+\beta)e^{-2}\{\phi_{j}(0) + \psi_{j}(0)\}],\label{e30}
\end{equation}

\begin{equation}
\rho_R = \frac{X}{\kappa R^2}\left[\Omega_{11}(X) + \Omega_{12}(X)
+ \beta \Omega_{13}(X) - 2\beta \Omega_{14}(X) - \frac{K}{2}
\Omega_{15}(X) + A\Omega_{16}(X)\right],\label{e30}
\end{equation}
where\\ $\phi_{j}(0) = \sum_{j=1}^{\infty}
\frac{(-1)^{j-1}}{j}$,\\ $\psi_{j}(0) =
\sum_{j=1}^{\infty}\frac{2^j}{j!j}$,\\
$\Omega_{11}(X)=(6+\beta)(1-X)^2(3-5X-4X^2)e^{2(X-1)}\{\phi_{j}(X)
+ \psi_{j}(X)\}]$,\\ $\Omega_{12}(X)=(6-11X^2+2X^3)$,\\
$\Omega_{13}(X)=\beta(-3+10X-7X^2)$,\\ $\Omega_{14}(X)=Xe^{2X-2}(1
- X)^2[1 + (1 - X)\theta_j(X)]- Ke^{2X}(1 -
X)^2(3-5X-4X^2)\Pi_{a,b,C,m,i,N}(X)$,\\
$\Omega_{15}(X)=(X)^{N+1}(1 - X)(a-bX)^m[4 e^{2X} + e^{2X}(1 -
X)]$,\\ $\Omega_{16}(X)= e^{2X}(-1 - 7X -7X^2-3X^3+4X^4)$.

\section{Physical Acceptability Conditions for Anisotropic Stellar Models}
In order to be physically meaningful, the interior solution for
static fluid spheres of Einstein's gravitational-field equations
must satisfy some general physical requirements. Because Einstein
field equation (2) high nonlinear in nature so not many realistic
physical solutions are known for the description of static
spherically symmetric perfect fluid spheres. Out of 127 solutions
only 16 were found to be physically meaningful
(\cite{Delgaty1998}). The following conditions have been generally
recognized to be crucial for anisotropic fluid spheres
\cite{Herrera1997}.

\subsection{Regularity and Reality Conditions}

\subsubsection{Case 1} The solution should be free from physical and geometrical
singularities i.e. pressure and energy density at the centre
should be finite and metric potentials $e^{-\lambda(r)}$ and
$e^{\nu(r)}$ should have non-zero positive values in the range $0
\leq r \leq R$. At origin Eq. (16) provides $e^{-\lambda(0)}=1$
whereas from Eq. (31) we obtain $e^{\nu(0)}=B$. So it is clear
that metric potentials are positive and finite at the centre (Fig.
1).

\begin{figure}[h]
\centering
\includegraphics[width=5cm]{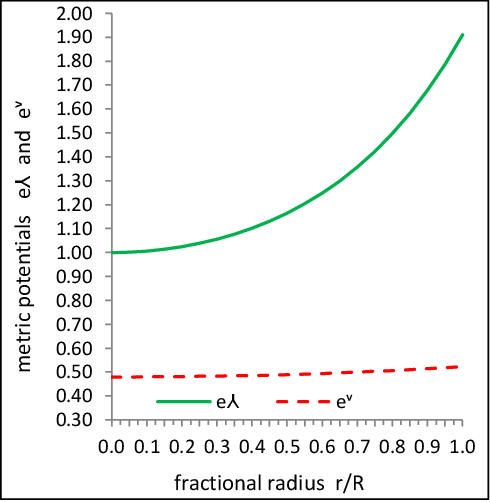}
\caption{Behavior of the metric potentials $\nu$ and $\lambda$ with
respect to fractional radial distance ($r/R$) for $RXJ 1856-37$. For plotting this figure, the numerical values of the parameters as follows: $n=1$,~$m=5$,~$N=11$,$a=1$,~$b=0.25$,~$K=0.11$,~$\beta=4.2395$, $A=1.1426$, $B=0.4798$, $C=2.3113 \times 10^{-3}$ (Table 5)}
\end{figure}

\subsubsection{Case 2} The density $\rho$ and radial pressure $p_r$ and tangential
pressure $p_t$ should be positive inside the star.

\begin{figure}[h]
\centering
\includegraphics[width=5cm]{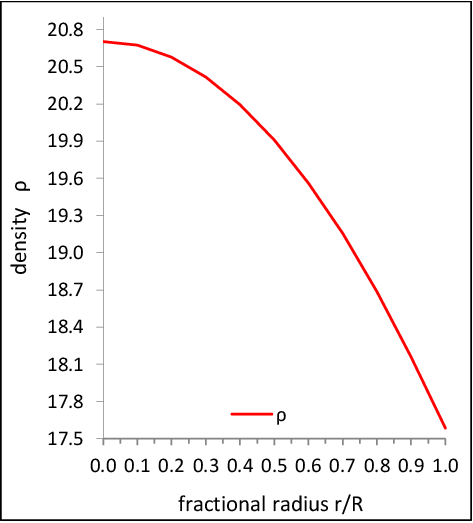}
\caption{Behavior of the effective matter-energy density $\rho_i=8\pi\,\rho/C$ with respect to
fractional radial distance ($r/R$) for $RXJ 1856-37$. For plotting of this figure we have employed
same data set of numerical values as used in Fig. 1}
\end{figure}

\begin{figure}[h]
\centering
\includegraphics[width=5cm]{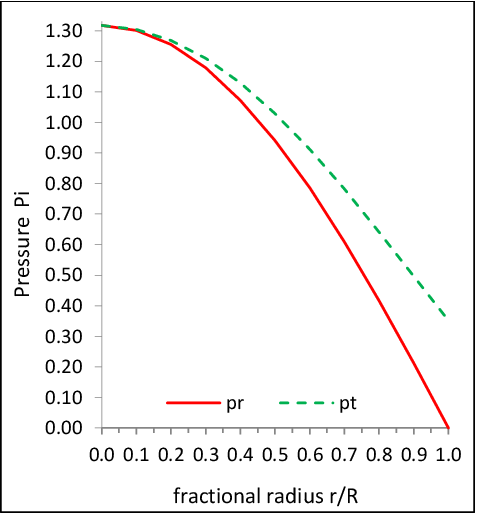}
\caption{Behavior of the effective radial and tangential pressures $P_r=8\pi\,p_r/C$ and $P_t=8\pi\,p_t/C$ with respect to the fractional radial distance $r/R$  for $RXJ 1856-37$. For plotting of this figure we have employed the same data set of numerical values as used in Fig. 1 and 2}
\end{figure}

\subsubsection{Case 3} The radial pressure $p_r$ must be vanishing at the boundary
of sphere $r=R$ but the tangential pressure $p_t$ may not vanish
at the boundary $r=R$ of the fluid sphere and may follow $p_t>0$
at $r=R$. However, the radial pressure is equal to the tangential
pressure at the centre of the fluid sphere.

\subsubsection{Case 4} $(dp_r/dr)_{r=0}= 0$ and $(d^2p_r/dr^2)_{r=0} < 0$ so that
pressure gradient $dp_r/dr$ is negative for $0 \leq r \leq R$.

\subsubsection{Case 5} $(dp_t/dr )_{r=0} = 0$ and $(d^2p_t/ dr^2)_{r=0} <
0$ so that pressure gradient $dp_t/dr$ is negative for $0 \leq r
\leq R$.

\subsubsection{Case 6} $(d{\rho}/dr )_{r=0} = 0$ and $(d^2\rho/dr^2)_{r=0} < 0$ so
that density gradient $d\rho/dr$  is negative for $0 \leq r \leq
R$.

Conditions (5.1.4) to (5.1.6) imply that pressure and density
should be maximum at the centre and monotonically decreasing
towards the surface (Figs. 2, 3).

\subsection{Causality and Well Behaved Conditions:}

\subsubsection{Case 1} Inside the fluid ball the speed of sound should be less than
the speed of light i.e. $0 \leq \sqrt{\frac{dp_r}{d\rho}} <1$, $0
\leq \sqrt{\frac{dp_t}{d\rho}} <1$ i.e. both
$\sqrt{\frac{dp_r}{d\rho}}$ and $\sqrt{\frac{dp_t}{d\rho}}$ are
lies between 0 and 1 which can be observed from Fig. 4 as well as
from Table 1.

\subsubsection{Case 2} The velocity of sound monotonically decreasing away from the
centre and it is increasing with the increase of density i.e.
$\frac{d}{dr}\left(\frac{dp_r}{d\rho}\right) <0$ or
$\frac{d^2p_r}{d\rho^2}>0$ and
$\frac{d}{dr}\left(\frac{dp_t}{d\rho}\right) <0$ or
$\frac{d^2p_t}{d\rho^2}>0$ for $0 \leq r \leq R$ (see Fig. 4). In
this context it is worth mentioning that the equation of state at
ultra-high distribution has the property that the sound speed is
decreasing outwards \cite{Canuto1973}.

\begin{figure}[h]
\centering
\includegraphics[width=5cm]{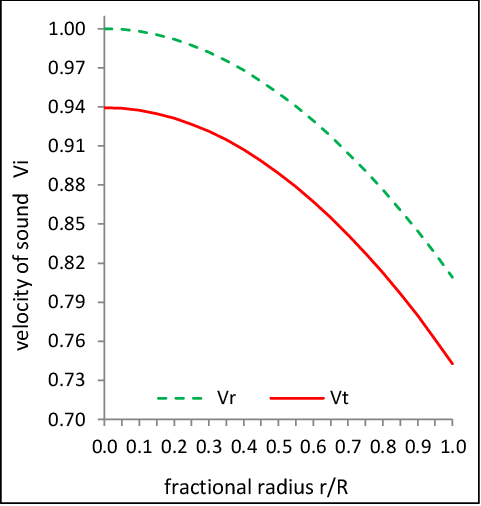}
\caption{Behavior of the sound speed $V$ with respect to the fractional radial distance ($r/R$)  for $RXJ 1856-37$. For plotting this figure, we have employed the same data set of numerical values as used in Figs. 1-3}
\end{figure}

\begin{table}[h]
\centering \caption{Values of different physical parameters $P_r=8\pi\,p_r/C$,\, $P_t=8\pi\,p_t/C$,\,
 $\rho_i=8\pi\,\rho/C$,\, $V_r=\sqrt{dp_r/d\rho}$,\, $V_t=\sqrt{dp_t/d\rho}$ \,for $RXJ 1856-37$ with the
 values of different parameters and constants: $n=1$,~$m=5$,~$N=11$, ~$a=1$,~$b=0.25$,~$K=0.11$,~$\beta=4.2395$.}\label{tbl-1}
\begin{tabular}{@{}lrrrrrrrrr@{}}
\hline

$r$     & $P_r$     & $P_t$  & $\rho_i$  & $V_r$  & $V_t$ \\ \hline

0.0     & 1.3165    & 1.3165 & 20.7045 & 0.9999 & 0.9394 \\

0.1     & 1.3009    & 1.3044 & 20.6724 & 0.9979 & 0.9374 \\

0.2     & 1.2543    & 1.2683 & 20.5763 & 0.9918 & 0.9312 \\

0.3     & 1.1776    & 1.2092 & 20.4164 & 0.9819 & 0.9211 \\

0.4     & 1.0723    & 1.1285 & 20.1933 & 0.9680 & 0.9070 \\

0.5     & 0.9406    & 1.0285 & 19.9076 & 0.9504 & 0.8890 \\

0.6     & 0.7852    & 0.9117 & 19.5603 & 0.9292 & 0.8672 \\

0.7     & 0.6093    & 0.7815 & 19.1526 & 0.9044 & 0.8417 \\

0.8     & 0.4167    & 0.6417 & 18.6860 & 0.8761 & 0.8125 \\

0.9     & 0.2120    & 0.4967 & 18.1624 & 0.8444 & 0.7795 \\

1.0     & 0.0000    & 0.3515 & 17.5839 & 0.8092 & 0.7427 \\ \hline
\end{tabular}
\end{table}

\subsubsection{Case 3} The ratios of the pressure to density, $p_r/\rho$ and
$p_t/\rho$ (as can easily be obtained from Table 1), should be
monotonically decreasing with the increase of $r$, i.e.
$\frac{d}{dr}\left(\frac{p_r}{\rho}\right)_{r=0}=0$ and
$\frac{d^2}{dr^2}\left(\frac{p_r}{\rho}\right)_{r=0}<0$,
$\frac{d}{dr}\left(\frac{p_t}{\rho}\right)_{r=0}=0$ and
$\frac{d^2}{dr^2}\left(\frac{p_t}{\rho}\right)_{r=0}<0$.
Then $\frac{d}{dr}\left(\frac{p_r}{\rho}\right)$ and
$\frac{d}{dr}\left(\frac{p_t}{\rho}\right)$ are negative valued
function for $r>0$. These behaviour can be observed from Fig. 5.
Also note from Table 1 which indicates the ratios via the data of $p_r$, $p_t$ and $\rho$.

\begin{figure}[h]
\centering
\includegraphics[width=5cm]{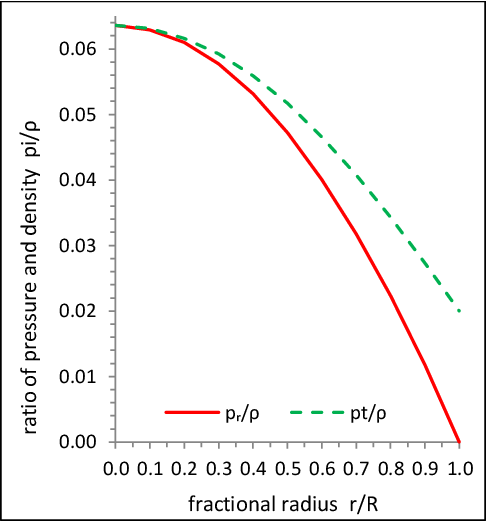}
\caption{Behavior of the ratios of pressure to density, $p_r/\rho$ and $p_t/\rho$ with respect to
the fractional radial distance ($r/R$)  for $RXJ 1856-37$. For plotting of this figure we have employed
same data set of numerical values as used in Fig. 1-4}
\end{figure}

\subsection{Energy Conditions}
A physically reasonable energy-momentum tensor has to obey the following energy conditions~\cite{Visser1996}:

\[(i)~NEC: \rho+E^2 \geq 0,\]

\[(ii)~WEC: \rho+p_r \geq 0,~\rho+p_t+2E^2 \geq 0,\]

\[(iii)~SEC: \rho+p_r+2p_t+2E^2 \geq 0.\]

\begin{figure}[h]
\centering
\includegraphics[width=5cm]{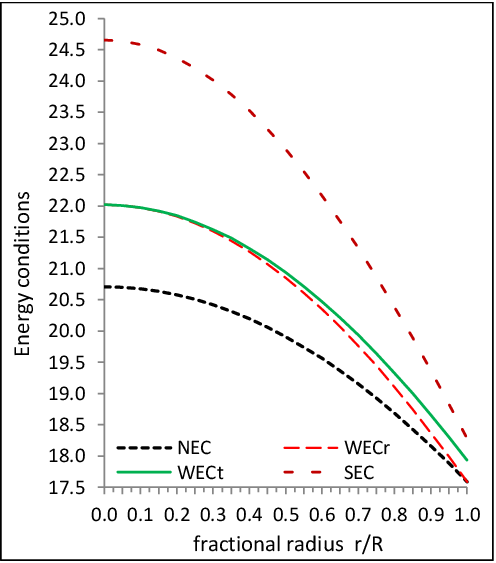}
\caption{Plot for the energy conditions with respect to the fractional radial distance
($r/R$) for $RXJ 1856-37$. For plotting of this figure we have employed same
data set of numerical values as used in Figs. 1-5}
\end{figure}

\begin{table}[h]
\centering \caption{Values of the energy conditions for the constants
$n=1$,~$m=5$,~$N=11$,\newline~$a=1$,~$b=0.25$,~$K=0.11$,~$\beta=4.2395$
and ~$CR^2 = 0.0829$ }\label{tbl-1}
\begin{tabular}{@{}lrrrrrrrrr@{}}
\hline $r/a$  & $NEC$  & $WEC_r$  & $WEC_t$  & $SEC$ \\ \hline
0.0 & 0.001861201 &0.001979549 & 0.001979549 & 0.002216246\\

0.1 & 0.001858755 & 0.001975725 & 0.001976041 & 0.002210297\\

0.2 & 0.001851417 & 0.001964274 & 0.001965539 & 0.002192519\\

0.3 & 0.001839186 & 0.001945266 & 0.001948116 & 0.002163126\\

0.4 & 0.001822059 & 0.001918816 & 0.001923889 & 0.002122476\\

0.5 & 0.001800038 & 0.001885088 & 0.001893032 & 0.002071076\\

0.6 & 0.001773127 & 0.001844301 & 0.001855771 & 0.002009588\\

0.7 & 0.001741340 & 0.001796733 & 0.001812391 & 0.001938834\\

0.8 & 0.001704704 &0.001742723 & 0.001763243 & 0.001859801\\

0.9 & 0.001663260 & 0.001682676 & 0.001708746 & 0.001773648\\

1.0 & 0.001617073 & 0.001617073 & 0.001649394 & 0.001681715 \\ \hline
\end{tabular}
\end{table}

Now we check whether all the energy conditions are satisfied or
not. For this purpose, numerical values of these energy conditions
are given in Table 2 and accordingly their behaviour are shown in
Fig. 6. This figure indicates that in our model all the energy
conditions are satisfied through out the interior region.

\subsection{Stability of the Stellar Models}

\subsubsection{Method 1} In order to have an equilibrium
configuration the matter must be stable against the collapse of
local regions. This requires, Le Chatelier's principle also known
as local or microscopic stability condition, that the radial
pressure $p_r$ must be a monotonically non-decreasing function of
$\rho$ \cite{Bayin1982}.

With the energy momentum tensor of the form (\ref{matter}), the
relativistic first law of   thermodynamics may be expressed as
\begin{equation}
\frac{d\rho}{p_r+\rho} = \frac{d\rho_m}{\rho_m}.\label{e43}
\end{equation}
where $p_r$ is the radial pressure, $\rho$ is the total energy
density and $\rho_m$ is that part of the mass density which
satisfies a continuity equation and is therefore conserved
throughout the motion.

We let the pressure change with density as
\begin{equation}
p_r \propto (\rho_m)^{\gamma}.\label{e44}
\end{equation}

From above Eq. (\ref{e44}) we have
\begin{equation}
\gamma = \left(\frac{\rho_m}{p_r}\right) \left(\frac{dp_r}{d\rho_m}\right).\label{e45}
\end{equation}

By Eqs. (\ref{e43}) and (\ref{e45}) we have
\begin{equation}
\gamma = \left(\frac{\rho+p_r}{p_r}\right) \left(\frac{dp_r}{d\rho}\right),\label{e46}
\end{equation}
where~$\gamma$~is a parameter called the adiabatic index. A
material obeying these equations is stable to gravitational
collapse if the pressure times the surface area increases more
rapidly than~$R^{-2}$. Because the density is proportional to
$R^{-3}$, the force exerted by the pressure is proportional to
$R^{2-3\gamma}$. This force increases more rapidly than the
gravitational force when $\gamma > 4/3$.

\begin{figure}[h]
\centering
\includegraphics[width=5.5cm]{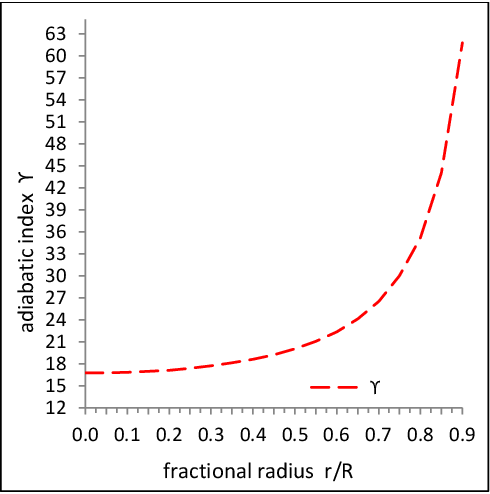}
\caption{Behaviour of the adiabatic index $\gamma$ with respect to the
fractional radial distance ($r/R$) for $RXJ 1856-37$. For plotting this figure,
we have employed the same data set of numerical values as used in Figs. 1-6}
\label{adiabatic}
\end{figure}

The later condition is, however, necessary but not sufficient to
obtain a dynamically stable model \cite{Tupper1983}. Heintzmann
and Hillebrandt \cite{Heintzmann1975} also proposed that neutron
star with anisotropic equation of state are stable for $\gamma >
4/3$.  Also it is well known that Newton's theory of gravitation
has no upper mass limit if the equation of state has an adiabatic
index $\gamma
> 4/3$.

The behavior of adiabatic index ($\gamma$) is shown in Fig.
\ref{adiabatic}. It is clear from figure that the value of
 $\gamma$ is more than $4/3$. So our model is stable.

\subsubsection{Method 2}

For this case let us write the generalized
Tolman-Oppenheimer-Volkoff (TOV) equation in the following form:
\begin{equation}
-\frac{M_G(\rho+p_r)}{r^2}e^{(\lambda-\nu)/2}-\frac{dp_r}{dr}+
\sigma \frac{q}{r^2}e^{\lambda/2} + \frac{2}{r}(p_t-p_r)=0,
\end{equation}
where $M_G$ is the gravitational mass within the radius $r$ and is
given by
\begin{equation}
M_G(r)=\frac{1}{2}r^2 \nu^{\prime}e^{(\nu - \lambda)/2}.
\end{equation}

Substituting the value of $M_G(r)$ in above equation we get
\begin{equation}
-\frac{1}{2} \nu^{\prime}(\rho+p_r)-\frac{dp_r}{dr}+ \sigma
\frac{q}{r^2}e^{\lambda/2}+\frac{2}{r}(p_t-p_r)=0.
\end{equation}

The above TOV equation describes the equilibrium condition for a
charged anisotropic fluid subject to gravitational ($F_g$),
hydrostatic ($F_h$), electric ($F_e$) and anisotropic stress
($F_a$) so that:
\begin{equation}
F_g+F_h+F_e+F_a=0,
\end{equation}
where
\begin{equation}
F_g=-\frac{1}{2} \nu^{\prime}(\rho+p_r),
\end{equation}

\begin{equation}
F_h=-\frac{dp_r}{dr},
\end{equation}

\begin{equation}
F_e=\sigma \frac{q}{r^2}e^{\lambda/2},
\end{equation}

\begin{equation}
F_a=\frac{2}{r}(p_t-p_r).
\end{equation}

Now, the above forces can be expressed in the explicit forms as
follows:
\begin{equation}
F_g= -\frac{1}{2} \nu^{\prime}(\rho+p_r)=-\frac{C
r}{1-x}(\rho+p_r),
\end{equation}

\begin{eqnarray}
F_h= -\frac{dp_r}{dr} =
-\frac{2C^2r}{8\pi}\left[A(1-4x+x^2+2x^3)e^{2x} + (10-14x+6x^2)+
Ke^{2x}(x)^N (1 - x)^3(a-bx)^m \right. \nonumber
\\ \left.
-2[(6+\beta)(1-x)(1+3x)-3(5-12x-3x^2+10x^3)]e^{2x-2}
(1+(1-x)\theta_j(x)) \right. \nonumber
\\ \left. - \beta[-2x+e^{2x-2}(-2+5x-3x^2)\{\phi_{j}(x) + \psi_{j}(x)\}]
\right. \nonumber
\\ \left.
-K\left(\frac{(1-x)^2(1-8x-11x^2-6x^3)e^{2x}\Pi_{a,b,C,m,i,N}(x)}{(1
+ x)^2} + \frac{\Psi_{a,b,C,m,i,N}(x)}{2}\right)\right],
\end{eqnarray}

\begin{equation}
F_e=\frac{K}{8\pi r^3}\left[e^{2x} (1-x)(x)^{N+2}(a-bx)^m
\left\{(N+3)(1-x)-2x+2(1-x)x -
\frac{bmx(1-x)}{(a-bx)}\right\}\right],
\end{equation}

\begin{equation}
F_a=\frac{2}{r}(p_t-p_r)=\frac{2\beta C}{8\pi r}x,
\end{equation}
with $x=Cr^2$ as mentioned earlier also.

\begin{figure}[h]
\centering
\includegraphics[width=5.5cm]{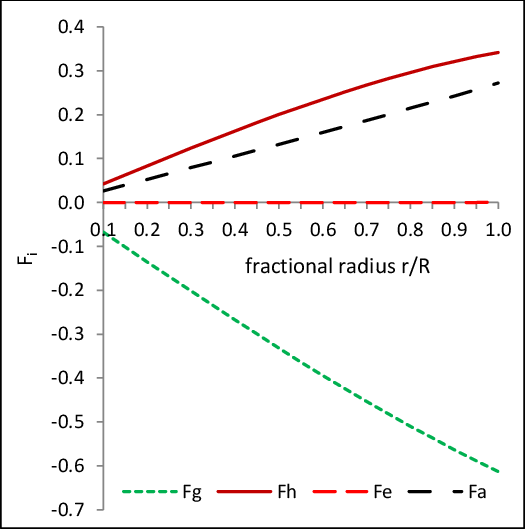}
\caption{Feature of the static equilibrium under different forces for $RXJ 1856-37$. For plotting this figure, we have employed the same data set of
numerical values as used in Figs. 1-7}
\label{forces}
\end{figure}

We have shown the plot for TOV equation in Fig. \ref{forces}. From
the figure it is observed that the system is in static equilibrium
under four different forces, e.g. gravitational, hydrostatic,
electric and anisotropic to attain overall equilibrium. However,
strong gravitational force is counter balanced jointly by
hydrostatic and anisotropic forces. The electric force seems has
negligible effect in this balancing mechanism.

\subsubsection{Method 3}

In our anisotropic model, to verify stability we plot the radial
($V_{sr}^2=dp_r/dt$) and transverse ($V_{st}^2=dp_t/d\rho$) sound
speeds in Fig. 9. It is observed that these parameters satisfy the
inequalities $0\leq V_{sr}^2 \leq 1$ and $0\leq V_{st}^2 \leq 1$
everywhere within the stellar object which obeys the anisotropic
fluid models \cite{Herrera1992,Abreu2007}.

\begin{figure}[h]
\centering
\includegraphics[width=5cm]{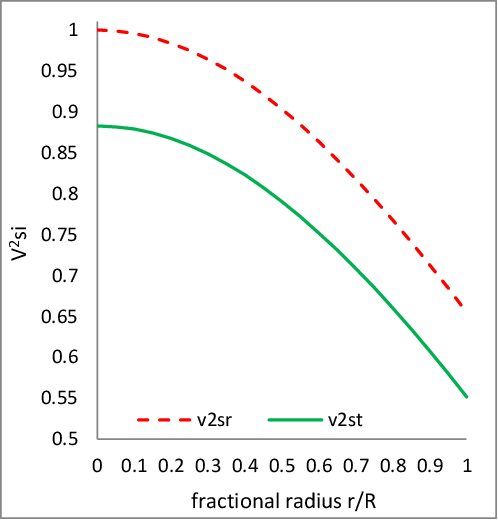}
\caption{Behavior of the square of the sound speed $V^2$ for $RXJ 1856-37$. For plotting this figure, we have employed the same data set of
numerical values as used in Figs. 1-8}
\end{figure}

\begin{figure}[h]
\centering
\includegraphics[width=5cm]{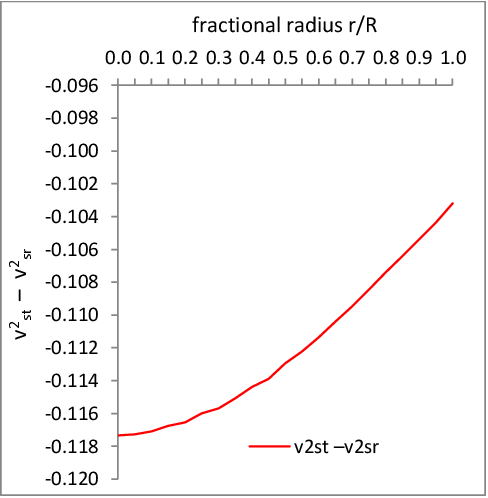}
\includegraphics[width=5cm]{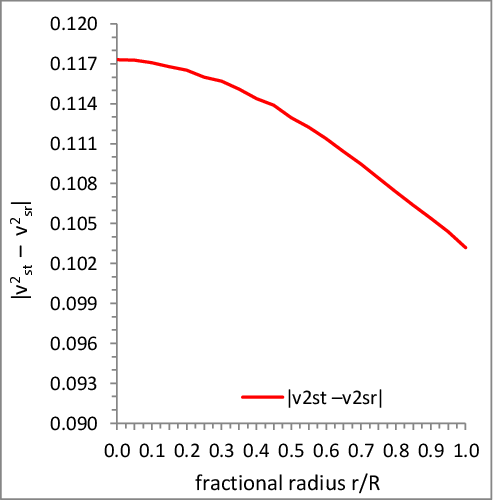}
\caption{Behavior of the difference of sound speeds
$V^2_{st}-V^2_{sr}$ and $\mid V_{st}^2 - V_{sr}^2 \mid$ for $RXJ 1856-37$.
For plotting of this figure we have employed same data set of numerical values as used in Fig. 1-9}
\end{figure}

Again, to check whether local anisotropic matter distribution is
stable or not, we use the proposal of Herrera \cite{Herrera1992},
known as cracking (or overturning) concept, which states that the
potentially stable region is that one where radial speed of sound
is greater than the transverse speed of sound. From the left panel
of Fig. 10, we can easily say that $V^2_{st}-V^2_{sr}\leq 1$.
Since, $0\leq V_{sr}^2 \leq 1$ and $0\leq V_{st}^2 \leq 1$,
therefore, $\mid V_{st}^2 - V_{sr}^2 \mid \leq 1 $ as can be seen
from the right panel of Fig 10. Hence, we can conclude that our
compact star model provides stable configuration.

\subsection{Electric charge}
From the present model it is observed that in the unit of Coulomb,
the charge on the boundary is $1.5151 \times 10^{13}$~C 
and at the centre it is as usual zero. In the Table 3 we have put
the data for charge $q$ in the relativistic unit Km. However, to
convert these values in Coulomb one has to multiply every value by
a factor $1.1659 \times 10^{20}$. Graphical plot is shown in Fig.
11 where charge profile is such that starting from a minimum it
acquires maximum value at the boundary.

\begin{table}[h]
\centering \caption{Values of charge, anisotropy and redshift for
$n=1$,~$m=5$,~$N=11$,\newline~$a=1$,~$b=0.25$,~$K=0.11$,~$\beta=4.2395$.
The related plots are shown \newline respectively in Figs. 11,~12
and 13}\label{tbl-1}
\begin{tabular}{@{}lrrrrrrrrr@{}}
\hline

$r$     & $q$ (Km)                &$\Delta_i$         & $Z$ \\
\hline

0.0     & 0                       & 0               &0.4437\\

0.1     & $1.3273 \times 10^{-21}$ &0.0035           & 0.4431 \\

0.2     & $2.1733 \times 10^{-17}$ &0.0141           & 0.4413\\

0.3     & $6.3374 \times 10^{-15}$  & 0.0316          & 0.4383\\

0.4     & $3.5511 \times 10^{-13}$  & 0.0562          & 0.4341\\

0.5     & $8.0574 \times 10^{-12}$  & 0.0879          & 0.4287\\

0.6     & $1.0318\times 10^{-10}$  & 0.1265          & 0.4220\\

0.7     & $8.9014 \times 10^{-10}$  & 0.1722          & 0.4141\\

0.8     & $5.7502 \times 10^{-9}$  & 0.2249          & 0.4049  \\

0.9     & $2.9775 \times 10^{-8}$  & 0.2847          & 0.3944  \\

1.0     & $1.2946 \times 10^{-7}$  & 0.3515          & 0.3826 \\
\hline
\end{tabular}
\end{table}

\begin{figure}[h]
\centering
\includegraphics[width=5cm]{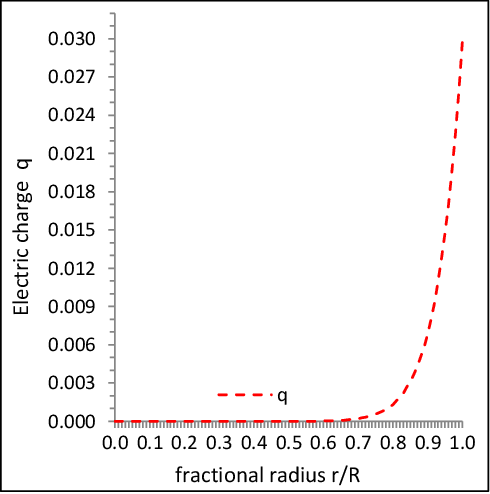}
\caption{Distribution of charge with respect to fractional radial distance ($r/R$) for $RXJ 1856-37$.
For plotting of this figure we have employed same data set of numerical values as used in Fig. 1-10}
\end{figure}

Let us now justify this feature of charge from the available
literature. It is shown by Varela et al. \cite{Varela2010} that
spheres with vanishing net charge contain fluid elements with
unbounded proper charge density located at the fluid-vacuum
interface and net charges can be huge ($10^{19}$~C). On the other
hand, Ray et al. \cite{SubharthiRay2003} have analyzed the effect
of charge in compact stars considering the limit of the maximum
amount of charge they can hold and shown through numerical
calculation that the global balance of the forces allows a huge
charge ($10^{20}$~Coulomb) to be present in a neutron star.
Thus we see that the net amount of charge has less effect to balance the
 mechanism of the force in our model.

\subsection{Pressure Anisotropy}

For the present model we calculate the measure of pressure
anisotropy as follows:
\begin{equation}
\Delta \equiv (p_t-p_r)= \beta C x. \label{eq28}
\end{equation}

It is in general argued that the `anisotropy' will be directed
outward for the condition $p_t > p_r$ i.e. $\Delta> 0 $, and
inward for the condition $p_t < p_r$ i.e. $\Delta < 0$. This
special feature can be observed from Fig. 12 related to our model.
This kind of repulsive `anisotropic' force allows for construction
of a more massive compact stellar configuration
\cite{Hossein2012}.

One can also calculate variation of the radial and transverse
pressures which are respectively given by $\frac{dp_r}{dr}$, as
can be obtained from Eq. (20), and $\frac{dp_t}{dr}=
\frac{dp_r}{dr} + \frac{2C^2{\beta} r}{8\pi}$.

\begin{figure}[h]
\centering
\includegraphics[width=5cm]{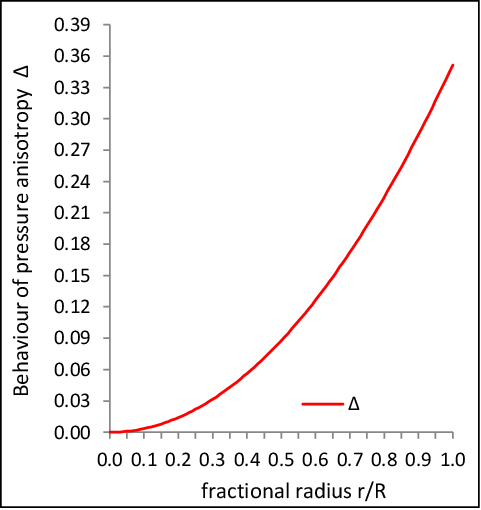}
\caption{Anisotropic behavior ($\Delta_i=\Delta/C$) at the stellar interior with
respect to fractional radial distance ($r/R$) for $RXJ 1856-37$.  For plotting of
this figure we have employed same data set of numerical values as used in Fig. 1-11}
\end{figure}

\subsection{Surface redshift}
The effective gravitational mass in terms of the energy density
can be written as
\begin{equation}
M_{eff}=4\pi\int_{0}^{R}\left(\rho +\frac{E^2}{8\pi}\right) r^2
dr= \frac{1}{2}R[1-e^{-\lambda(R)}], \label{eq33}
\end{equation}
where $e^{-\lambda(R)}$ is given by Eq. (46).

One can therefore provide the compactness of the star as
\begin{equation}
u= \frac{M_{eff}}{R}= \frac{1}{2}[1-e^{-\lambda(R)}] \label{eq35}
\end{equation}

Again we define the surface redshift corresponding to the above
compactness factor as follows:
\begin{equation}
Z = [1-2u]^{-1/2} - 1 = e^{\lambda(R)/2} - 1. \label{eq36}
\end{equation}

\begin{figure}[h]
\centering
\includegraphics[width=5cm]{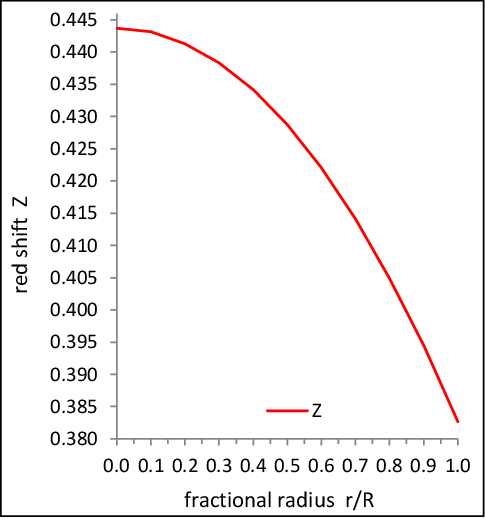}
\caption{Redshift of the stellar model with respect to
fractional radial distance ($r/R$) for $RXJ 1856-37$. For plotting this figure, we have employed the same
data set of numerical values as used in Figs. 1-12}
\end{figure}

We plot redshift in Fig. 13 from which it is evident that
it is showing a gradual decrease. This feature also can be
observed from the Table 3. The maximum surface redshift for the
present stellar configuration of radius $6.0$~km turns out to be
$Z = 0.3826$.

In this connection it is to mention that for isotropic case and in
the absence of the cosmological constant the surface redshift is
constraint as $Z \leq
2$~\cite{Buchdahl1959,Straumann1984,Boehmer2007}. Again for an
anisotropic star in the presence of a cosmological constant the
constraint on surface redshift is $Z \leq 5$ \cite{Boehmer2006}
whereas Ivanov \cite{Ivanov2002} put the bound $Z \leq 5.211$.
Based on the above discussion we therefore conclude that for an
anisotropic star without cosmological constant the value for our
model $Z = 0.3826$ is in good agreement.

\section{Some Case Studies: Comparison of Present Stellar Model with Compact Stars}

\subsection{Allowable Mass to Radius Ratio}

Buchdahl \cite{Buchdahl1959} has proposed an absolute constraint
of the maximally allowable mass-to-radius ratio $(M/R)$ for
isotropic fluid spheres of the form $2M/R \leq 8/9$ (in the unit,
$c = G = 1$) which states that for a given radius a static
isotropic fluid sphere cannot be arbitrarily massive. B{\"o}hmer
and Harko \cite{Boehmer2007} proved that for a compact object with
charge, $Q (<M)$, there is a lower bound for the mass-radius
ratio
\begin{equation}
\frac{3Q^2}{2R^2}
\left(\frac{1+\frac{Q^2}{18R^2}}{1+\frac{Q^2}{12R^2}}\right) \leq
\frac{2M}{R}.\label{e47}
\end{equation}

Upper bound of the mass of charged sphere was generalized by
Andr{\'e}asson \cite{Andreasson2009} and proved that
\begin{equation}
\sqrt{M} \leq \frac{\sqrt{R}}{3} + \sqrt{\frac{R}{9} +
\frac{Q^2}{3R}}.\label{e48}
\end{equation}

\begin{figure}[h]
\centering
\includegraphics[width=5cm]{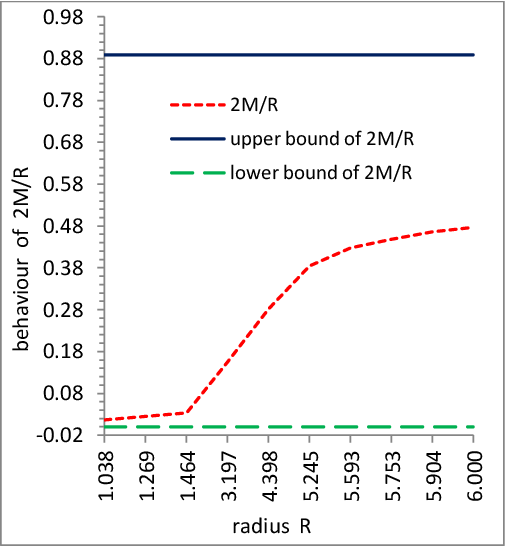}
\caption{Plot for mass to radius ratio $M/R$ with respect to
radius $R$ for $RXJ 1856-37$}
\end{figure}

By substituting the following data, mass $M= 0.9693$ Solar mass
and radius $R=6.0$ Km, we find out that $M/R=0.238 < 4/9$ and also
$2M/R=0.4760$ which satisfy Buchdahl condition of stable
configuration \cite{Buchdahl1959}. We also note from Fig. 14 and
Table 4 that the charged stars have large mass and radius as we
should expect due to the effect of the repulsive Coulomb force
with the $M/R$ ratio increasing with charge
\cite{SubharthiRay2003}. However, unlike Ray et al.
\cite{SubharthiRay2003} where in the limit of the maximum charge
the mass goes up to $10$, which is much higher than the maximum
mass allowed for a neutral compact star, our model seems very
satisfactory.

\begin{table}[h]
\centering \caption{Values of lower and upper bounds of mass for
the constants
$n=1$,~$m=5$,~$N=11$,\newline~$a=1$,~$b=0.25$,~$K=0.11$ and
$\beta=4.2395$}\label{tbl-1}
\begin{tabular}{@{}lrrrrrrrrr@{}}
\hline $R$  & $2M/R$  & $Lower Bound$  & $Upper Bound$ \\ \hline

0.7350 & 0.0085 & $8.2364 \times 10^{-31}$ &  0.8889\\

1.0379 & 0.0169 & $6.7464 \times 10^{-27}$ & 0.8889\\

1.2693 & 0.0252 & $1.3147 \times 10^{-24}$ & 0.8889\\

1.4635 & 0.0334 & $5.5491 \times 10^{-23}$ &  0.8889\\

3.1974 & 0.1549 & $8.0491 \times 10^{-14}$ & 0.8889\\

4.3975 & 0.2819 & $9.8941 \times 10^{-10}$ & 0.8889\\

5.2445 & 0.3847 & $3.1413 \times 10^{-7}$ & 0.8889\\

5.5925 & 0.4280 & $2.9932 \times 10^{-6}$ & 0.8889\\

5.7525 & 0.4479 & $8.3841 \times 10^{-6}$ & 0.8889\\

5.9043 & 0.4665 & $2.1783 \times 10^{-5}$ & 0.8889\\

6.0000 & 0.4760 & $3.7139 \times 10^{-5}$ & 0.8889 \\ \hline
\end{tabular}
\end{table}

\subsection{Validity with Strange Star Candidates}
We have presented two tables here (Tables 5 and 6) from where it
can be observed that the mass and radius are exactly correspond to
the strange stars $RXJ~1856-37$ and $Her~X-1$. What we did in the
tables are as follows: by considering the mass and radius of the
above mentioned stars we have figured out data for the model
parameters, and in the next step we evaluated data for different
physical parameters, e.g. central density, surface density and
central pressure, of those strange stars. One can observe that
these data set are in good agreement with the available
observational data.

In this connection we would like to mention that previously
Gupta and Maurya~\cite{Gupta2011} showed a similar result for
$PSR~J~1614-2230$ with isotropic fluid distribution and charge
generalization of Durgapal~\cite{Durgapal1985}. We also note that
like the models offered by Kalam et al. \cite{Kalam2012}, Hossein
et al. \cite{Hossein2012} and Kalam et al. \cite{Kalam2013} our
presented models provide significantly promising results with
observational evidences.

\begin{table}[h]
\centering \caption{Values of the model parameters $A$, $B$, $m$
etc. for different strange stars}\label{tbl-1}
\begin{tabular}{@{}lrrrrrrrrrrrrr@{}} \hline
Strange star    &$M$    & $R$   & $A$  & $B$  & $m$ &$N$ & $a$ &
$b$ & $K$ & $\beta$ & $C(km^{-2})$ \\ candidates & ($M_{\odot}$) & (Km)\\
\hline

$RX~J~1856-37$  & 0.9693    & 6.0   & 1.1426    & 0.4798 &5 &11 &1
                &0.25       &0.11   &4.2395     &$2.3113 \times 10^{-3}$\\

$Her~X-1$       &0.88       &7.7    &1.7518 &0.6217 &4 &14 &1
                &0.16       &0.13   &2.1546     &$1.0760 \times 10^{-3}$\\ \hline
\end{tabular}
\end{table}

\begin{table}[h]
\centering \caption{Energy densities and pressure for different
strange star candidates \newline for the above parameter values of
Table 1}\label{tbl-2}
\begin{tabular}{@{}lrrrrrrrrr@{}}
\hline

Strange star        & Central Density           & Surface density
& Central pressure \\

candidates          & ($gm/cm^3$)            & ($gm/cm^3$) &
($dyne/cm^2$)\\ \hline

$RXJ~1856-37$       & $2.5119 \times 10^{15}$   & $2.1824  \times
10^{15}$ & $1.4378  \times 10^{35}$\\

$Her~X-1$           & $1.0925  \times 10^{15}$  & $1.0116 \times
10^{15}$ & $6.0308 \times 10^{34}$\\ \hline
\end{tabular}
\end{table}

\section{Conclusion}

In this work we have presented a set of new solutions for an
anisotropic charged fluid distribution under the framework of
General Theory of Relativity. To solve the Einstein-Maxwell field
equations we construct a general algorithm for all possible
anisotropic charged fluid spheres. As an additional condition
which simplifies the physical system of space-time we consider a
special source function in terms of metric potential $\nu$. We
further adopt exterior solution of Reissner-Nordstr{\"o}m so that
our interior solution can be matched smoothly as a consequence of
junction conditions at the surface of spheres $(r=R)$.

The solutions set thus obtained exhibits regular physical
behaviour as can be observed from figures and tables on different
parameters. We specifically discuss (i) regularity and reality
conditions (applied for metric potentials $e^{-\lambda(r)}$ and
$e^{\nu(r)}$, energy density $\rho$, fluid pressures $p_r$ and
$p_t$, pressure gradients $dp_r/dr$ and $dp_t/dr$, and density
gradient $d\rho/dr$), and (ii) causality and well behaved
conditions (applied for speed of sound $dp_r/d\rho$ and ratios of
pressure to densities $p_r/\rho$ and $p_t/\rho$). Beside all these
general physical properties the solutions set shows desirable and
essential features for energy condition, stability condition,
charge distribution, pressure anisotropy and surface redshift.
Among these physical parameters as a special case, regarding
electric charge distribution of our model, we note that the charge
on the boundary is $1.5151 \times 10^{13}$~Coulomb and at the
centre it is as usual zero. Other features of charge is also
available in the literature
\cite{SubharthiRay2003,Varela2010,Murad2014} in connection to
stable configuration of compact stars where it has been shown that
the global balance of the forces allows a huge charge ($\sim
10^{20}$~Coulomb) to be present in a neutron star.

We also observe some special and interesting features for our
stellar models which are related to compact stars as follows:

(1) Allowable mass to radius ratio: The condition of Buchdahl
\cite{Buchdahl1959} related to the maximally allowable
mass-to-radius ratio for isotropic fluid spheres is of the form
$2M/R \leq 8/9$. By substituting the following data, mass $M=
0.9693~M_{\odot}$ and radius $R=6.0$~Km, we find out that
$2M/R=0.4760$ which satisfies Buchdahl condition of stable
configuration \cite{Buchdahl1959} as mentioned above.

(2) Validity with strange stars: We have prepared several data set
from where it is observed that the mass and radius are exactly
correspond to the strange stars $RXJ~1856-37$ and $Her~X-1$.
Therefore, one can note that like the models of Kalam et al.
\cite{Kalam2012}, Hossein et al. \cite{Hossein2012} and Kalam et
al. \cite{Kalam2013} our models also provide significantly
promising results with observational evidences.

In this work we have studied the case for $n=1$ only in the source
function because of the fact that this value is more relevant for
exploring existence and properties of strange stars. There is
however scope for further study with other values of $n$ also as
follows: (1) For integer values of $n=2,~3,~5$ (not possible
for all other positive integer values), and (2) For fractional
values of $n$ there are two possibilities: (i) If $n$ lies between
$0$ and $1$ then exact solutions are possible for all fractional
values, and (ii) If $n$ is greater than $1$ then for all
fractional values of $n$ except the values of $n=p/(p-1)$ and
$n=p/(p-2)$, where $p$ is a positive integer ($p \neq 1$ and $p
\neq 2$). However, the specific value $3/2$ is not allowed for
these factors to study the solutions for the present model.

As a final comment we would like to mention that Tiwari and Ray
\cite{Tiwari1991} proved that any relativistic solution for
spherically symmetric charged fluid sphere has electromagnetic
origin and hence provides {\it Electromagnetic Mass} model
\cite{Lorentz1904,Wheeler1962,Feynman1964,Wilzchek1999}.
Therefore, it would be an interesting task to verify whether our
model also represents an electromagnetic mass or not and can be
studied elsewhere in a future project.

\section*{Appendix}

\begin{table}[h]
\centering \caption{List of regular behavior of $dp/d\rho$ for the
{\it ansatz}~$e^{\nu(r)} = B(1+x)^n$,\newline~with $x=Cr^2$ where
$p$ in the table is a positive integer}\label{tbl-4}
\begin{tabular}{@{}lrrrrrrrrr@{}}
\hline
 n   & Electric charge function   & Pressure anisotropy & Behavior of $dp/d\rho$ & Reference\\
      & ($2q^2 C/x^2$)                &                            &                                 &\\
\hline

 1   & 0     & 0   & No & ~\cite{Tolman1939}\\

 1    & Kx   & 0   & Yes & ~\cite{PR2011}\\

2     & 0 & 0   & No & ~\cite{Wyman1949,Adler1974}\\

1,2,7  & $Kx[1+(n+1)x]^{\frac{n-1}{n+1}}$ & 0 & Yes & ~\cite{PR2011}\\

2   & $Kx(1+x)^N[1+(n+1)x]^{m+\frac{n-1}{n+1}},~N \neq 2$ & 0 &  Yes & ~\cite{FM2013}\\

2   & $Kx^{N+1}(1+x)^{1-N}$ & 0  & Yes &  ~\cite{MF2013}\\

2   & $Kx^{N+1}(1+mx)^{p}[1+(n+1)x]^{\frac{n-1}{n+1}}$  & 0  & Yes & ~\cite{MF2013}\\

2   & $Kx^{N+1}(1+mx)^{p}(1+x)^{1-n}[1+(n+1)x]^{\frac{n-1}{n+1}}$  & 0 & Yes &  ~\cite{MF2013}\\

2   & $Kx^{N+1}(1+x)^{1-n}(1+mx)^{p}[1+(n+1)x]^{\frac{n-1}{n+1}}$ & $\delta x(1-2ax)^{1-n}$ & Yes & ~\cite{Murad2014}\\

     &   & $\times [1+(n+1)x]^{\frac{n-1}{n+1}}$ & \\

3  & $Kx(1+x)^{n}(1+4x)^{1/2}$ & 0  & Yes & ~\cite{PN2012}\\

4  & $Kx^{n}(1+x)^{-2}$ & 0  & Yes & ~\cite{MGP2011}\\

5  & $Kx(1+6x)^{2/3}$ & 0  & Yes & ~\cite{GM2011a}\\

6  & $Kx(1+7x)^{5/7}$ & 0  & Yes & ~\cite{MG2011b}\\

n  & 0 & 0  & Yes~($n \geq 4$) & ~\cite{Maurya2011}\\

n  & $n^2Kx[1+(n+1)x]^{\frac{n-1}{n+1}}$ & 0  & Yes~($n \geq 1$) & ~\cite{MG2011c}\\

n  & 0 & $n^2C\Delta_{0}x$ & Yes~($n \geq 4$) & ~\cite{MG2012}\\

    &    & $\times [1+(n+1)x]^{\frac{n-1}{n+1}}$ & \\

n  & $n^2Kx[1+(n+1)x]^{\frac{n-1}{n+1}}$ & $n^2C\Delta_{0}x$     & Yes  & ~\cite{MG2014}\\
    &  & $\times [1+(n+1)x]^{\frac{n-1}{n+1}}$ & \\ \hline
\end{tabular}
\end{table}

\begin{table}[h]
\centering
\caption{List of regular behavior of $dp/d\rho$ for the
{\it ansatz} $e^{\nu(r)} = B(1-x)^{-n}$}\label{tbl-4}
\begin{tabular}{@{}lrrrrrrrrr@{}}
\hline
 n   & Electric charge function   & Pressure anisotropy & Behavior of $dp/d\rho$ & Reference\\
      & ($2q^2 C/x^2$)                &                            &                                 &\\
\hline
$0<n<1$   & 0  & 0  & Yes~($N \geq 10$),\\

     &   &   & ~$N=\frac{1+n}{1-n}$,~$N \in I^{+}$,~$N>1$ & ~\cite{Maurya2012c}\\

$1/3$   & $Kx(3-2x)^{-2}$ & 0 & Yes & ~\cite{Maurya2012b}\\

$0<n<1$  & $Kx(1-\{2/[N+1]\}x)^{-N}$, & 0 &Yes~($N \geq 2$) & ~\cite{Maurya2012a}\\

    &~$N=\frac{1+n}{1-n}$,~$N \in I^{+}$,~$N>1$  & \\

$0<n<1$  & $2E^2=\Delta$  & $C\Delta_{0}x[1-(1-n)x]^{-N}$ & Yes~($N \geq 2$),\\

    &  &    & $N=\frac{1+n}{1-n}$,~$N \in I^{+}$ &  ~\cite{Maurya2013}\\

1,2,3  & 0 & 0 & Yes~($N=1,3$) & ~\cite{Maurya2014b}\\

1,2,3  & $\frac{K}{2}x^n[1+(n-1)x]^{n-1}$ & 0 & Yes & ~\cite{Maurya2014a}\\

1,n    & $Kx^{1+N}(1-x)^{n+1}(a-bx)^m$ & $\beta x^{n}[1+(n-1)x]^{n-1}$ & Yes & [Present paper]\\

& $\times [1+(n-1)x]^{\frac{n-1}{n+1}}$ & &\\ \hline
\end{tabular}
\end{table}

\section*{Acknowledgement}
SKM acknowledges support from the Authority of University of
Nizwa, Nizwa, Sultanate of Oman. Also the author SR is thankful to
the authority of Inter-University Centre for Astronomy and
Astrophysics, Pune, India for providing him Associateship
programme under which a part of this work was carried out.

\end{document}